# Sensitivity of Interpolated Yield Optimized Superoscillations

Ehud Perlsman and Moshe Schwartz

School of Physics and Astronomy , Raymond and Beverly Sackler Faculty of Exact Sciences,

Tel Aviv University, Tel Aviv 69978, Israel

Abstract- Super oscillating signals are band limited signals that oscillate in some region faster than their largest Fourier component. Such signals have many obvious scientific and technological applications, yet their practical use is strongly limited by the fact that an overwhelming proportion of the energy goes into that part of the signal, which is not superoscillating. In a recent article the problem of optimization of such signals has been studied. In that article the concept of superoscillation yield is defined as the ratio of the energy in the super oscillations to the total energy of the signal, given the range in time and frequency of the superoscillations, which is imposed by forcing the signal to interpolate among a set of predetermined points. The optimization of the superoscillation yield consists of obtaining the Fourier coefficients of the low frequency components of which the signal consists, that maximize the yield under the interpolation constraint. Since in practical applications it is impossible to determine the Fourier coefficients with infinite precision, it is necessary to answer two questions. The first is how is the superoscillating nature of the signal affected by random small deviations in those Fourier coefficients and the second is how is the yield affected? These are the questions addressed in the present article. Limits on the necessary precision are obtained. Those limits seem not to be impractical.

Super oscillatory functions are band limited functions that oscillate over some intervals with a frequency larger than its maximal Fourier component. A number of examples have been given in the past for such functions with very interesting applications to Quantum Mechanics [1]-[6] , signal processing [7]-[11] and to optics , where super oscillations are intimately related to super resolution [12]-[14] . Although the phenomenon of superoscillation suggest many practical applications, it's uses are presently rather limited, because they exist in limited intervals and that the amplitude of the superoscillation in those regions is extremely small compared to typical values of the amplitude in non-oscillating regions [2],[10],[11] . The Aharonov functions

$$f_n(x) = [\cos(x/n) + ic\sin(x/n)]^n , \quad \text{with } c > 1 , \qquad (1)$$

provide a generic example. When the function is decomposed into its Fourier components it is evident that the highest frequency is 1. Yet for values of $x$ of the order of $\sqrt{n}$, very small compared to $2\pi n$, the periodicity of the function, it behaves as $e^{icx}$. By considering

$$|f_n(x)|^2 = [1+(c^2-1)\sin^2(x/n)]^n , \qquad (2)$$

it becomes clear when $x$ is of order $n$ the expression in the square brackets is typically larger than one by a finite, $n$ independent amount, and thus $|f_n(x)|^2$ is exponentially large in $n$, while in the superoscillating interval it is close to one.

For the benefit of the reader we repeat here shortly the construction of interpolated superoscillations and the definition of superosillation yield. Consider the function

$$f(x) = \frac{A_0}{\sqrt{2\pi}} + \sum_{m=1}^{N} \frac{A_m}{\sqrt{\pi}} \cos(mx) , \qquad (3)$$

Choose an interval [0,a] with $a < \pi$ and impose on the function $M$ constraints in the interval, $f(aj/M-1) = (-1)^j \equiv \mu_j$ for $j=0,...,M-1$[11,15]. The constraints result in a set of $M$ linear equations in $N+1$ unknowns of the form,

$$\sum_{m=0}^{N} C_{jm} A_m \equiv \mathbf{C}_j \cdot \mathbf{A} = \mu_j , \qquad (4)$$

where $C_{jm} = \frac{1}{\sqrt{\pi}} \begin{cases} \cos(ajm/M) & \text{for } m \neq 0 \\ 1/\sqrt{2} & \text{for } m = 0. \end{cases}$

Provided $M \leq N+1$, this choice constrains the function to oscillate within the interval $[-a,a]$ between the values $\pm 1$ with a frequency

$$\omega = \frac{\pi(M-1)}{a} . \qquad (5)$$

The superoscillation yield has been defined in ref. 15 as the ratio between the energy in the superoscillation interval and the total energy per period

$$Y(M,a) = \frac{\int_{-a}^{a} f^2(t)dt}{\int_{-\infty}^{\infty} f^2(t)dt}.  \tag{6}$$

The yield is maximized with respect to the choice of the $A_m$'s under the constraints given by equation (4). The result is a set $\{\tilde{A}_m\}$ of optimal Fourier coefficients. For practical uses two questions have to be considered, resulting from the finite precision that can be expected when the Fourier coefficients are determined. The first is to what extent superoscillation with the original frequency still persists under random fluctuations around the optimal Fourier coefficients and the second is what happens to the superoscillation yield under such fluctuations. To address those questions assume that the actual Fourier coefficients describing the superoscillating function are given by

$$\hat{A}_m = \tilde{A}_m + \delta_m,  \tag{7}$$

where $\delta_m$ is random and governed by a Gaussian distribution with standard deviation $\delta$. Consider the first question, which is the sensitivity of the interpolation to random fluctuations around the optimal solution . Equation (4) implies that the change in the value of the function $f(x)$ at the points $x_j = ja/M$ for $j=0,...,M-1$, is given by

$$\delta f_j = C_{jm} \delta_m,  \tag{8}$$

where all the absolute values of all the $C_{jm}$'s are lessor equal 1. Consequently,

$$\left|\delta f_j\right| \propto \delta \sqrt{N}.  \tag{9}$$

Now as long as the right hand side above is (considerably) less than one,

$$sign[f(x_j)] = (-1)^j,  \tag{10}$$

So that the number of oscillations in the interval $[0,a]$ does not decrease and thus the superoscillation is still preserved. Note, however, that the accuracy required in the determination of the Fourier coefficients is not relative but absolute. The optimal Fourier coefficients are exponentially large in $N$ and this is why out of the superoscillation interval the absolute value of $f$ is exponentially large in $N$. ( The fact that within the superoscillation interval the function is of order 1 is due to cancellations, resulting partly from the strong fluctuations in sign of the Fourier coefficients.) The relative accuracy required thus in the Fourier coefficients is of the order $\theta = \delta/|A| \propto N^{-1/2} \alpha^{-N}$, where $|A|$ denotes the typical size of the $\tilde{A}_m$'s and $\alpha > 1$. Thus, if it is only relative accuracy that is possible to achieve, care should be taken that $N$ is not too large to preserve the superoscillations. If absolute accuracy can be obtained, regardless of the size of the optimal Fourier coefficients, it should be no problem to preserve superoscillation under very reasonable fluctuations in those quantities. We characterize the departure from the enforced interpolation, for a given set of errors in the Fourier coefficients by a single parameter,

$$\delta V = \frac{1}{M} \sum_{i=0}^{M-1} [\delta f_i]^2 \ . \tag{11}$$

The following graphs give the interpolation sensitivity, which is the average of $\delta V$ over the Gaussian distribution of the $\delta_m$'s as a function of $\delta$ for two examples of yield optimized superoscillations,

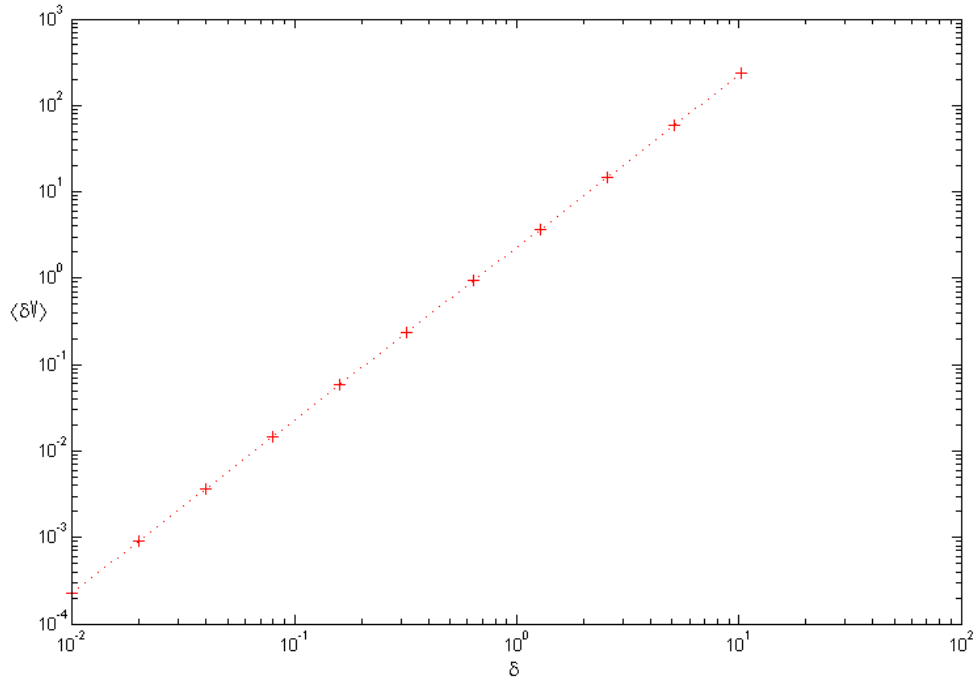

Fig.1- The interpolation sensitivity as a function of $\delta$ for $N = 6$, $M = 4$

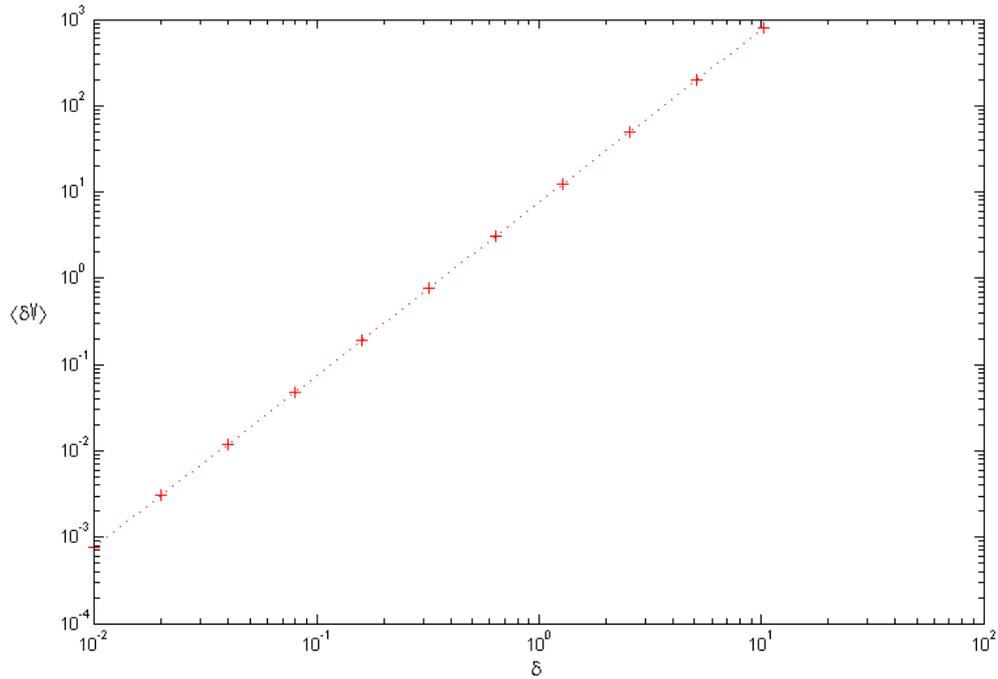

Fig.2- The interpolation sensitivity as a function of $\delta$ for $N=12$, $M=8$

Note that in spite of the fact that in the graphs the $\delta$ attain values up to $10^2$, only for $\delta$'s considerably smaller than one, we may expect the original frequency in the superoscillating interval to be unchanged. (The oscillation period, $X$, is taken as the average peak to peak distance and the frequency is $\omega = 2\pi/X$.)

While the interpolation sensitivity, $\langle \delta V \rangle$ is an important quantity that characterizes the deviations from the enforced interpolation, it is, obviously, not less important to obtain the distribution of the quantity $\delta V$. In the following we give that distribution for the case $N=12$, $M=8$ and for a specific value of $\delta = 0.01$. It is clear that at least as far as the order of magnitude is considered, the interpolation sensitivity represents well the the distribution of the values of $\delta V$. The distribution is obtained by consideration of $10^7$ realizations of the set of the low frequency Fourier coefficients.

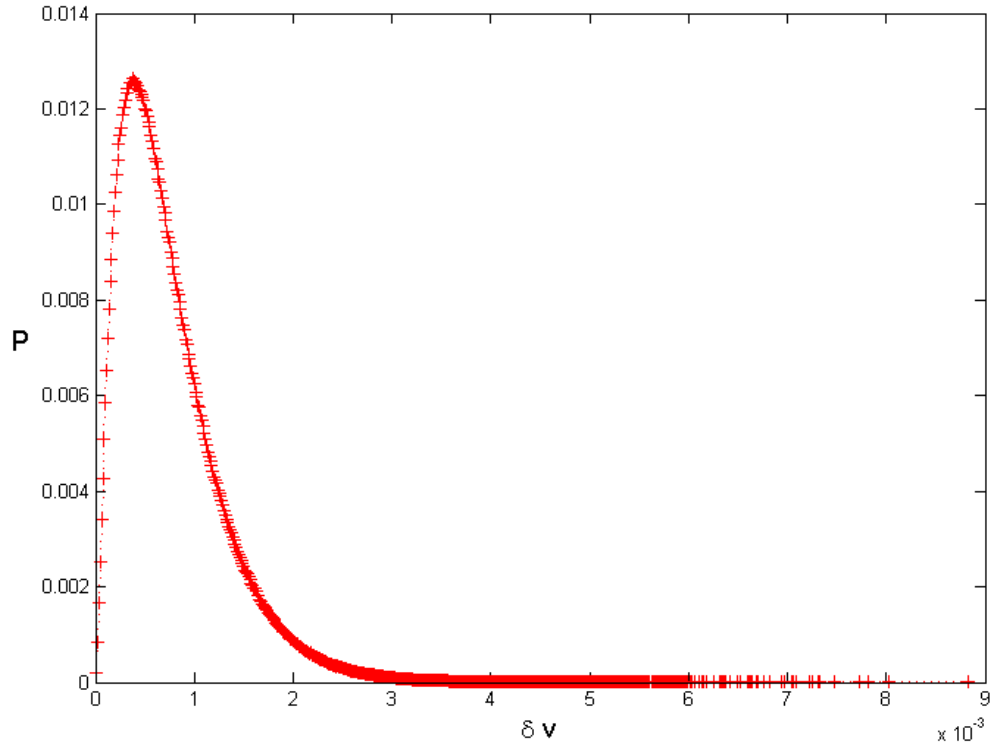

Fig. 3 – The normalized distribution, P, of $\delta V$, for $\delta = 0.01$. The distribution is obtained from $10^7$ realizations of the set of Fourier coefficients.

The next problem is that of the sensitivity of the superoscillation yield. The yield can be expressed in terms of the deviations as follows

The deviations from the optimal Fourier coefficients define a function giving the deviation from the optimal function, $\tilde{f}(x)$, obeying the constraints. That deviation is given by

$$\delta(x) = \frac{\delta_0}{\sqrt{2\pi}} + \sum_{m=1}^{N} \frac{\delta_m}{\sqrt{\pi}} \cos(mx). \qquad (11)$$

It is thus clear that a bound on the average of the square of the deviation function is

$$\langle \delta^2(x) \rangle < \frac{N}{\pi^2} \delta^2 \qquad (12)$$

regardless of the value of $x$. The first order change of the yield is

$$\delta Y = 2 \frac{\int_{-a}^{a} \tilde{f}(x)\delta(x)dx}{\int_{-\pi}^{\pi} \tilde{f}^2(x)dx} - 2Y \frac{\int_{-\pi}^{\pi} \tilde{f}(x)\delta(x)dx}{\int_{-\pi}^{\pi} \tilde{f}^2(x)dx} \quad . \tag{13}$$

The fact that the yield has a linear correction when calculated at its maximum, should not be surprising, because it is a constrained maximum and the linear part appears as a result of the small violations of the strict interpolation constraint. Had we considered deviations in the Fourier coefficients that honor the constraints, the lowest order correction would be quadratic in $\delta(x)$. This implies, that because of the departure from the interpolation constraints, the yield can increase under fluctuations. In the next two figures we give the average of $\delta Y$, as a function of $\delta$. As we shall see that average, which is not the average of the linear approximation of $\delta Y$ given above, is positive. This will be explained when the scatter graph, giving the distribution of pairs $(\delta V, \delta Y)$ for given $\delta$.

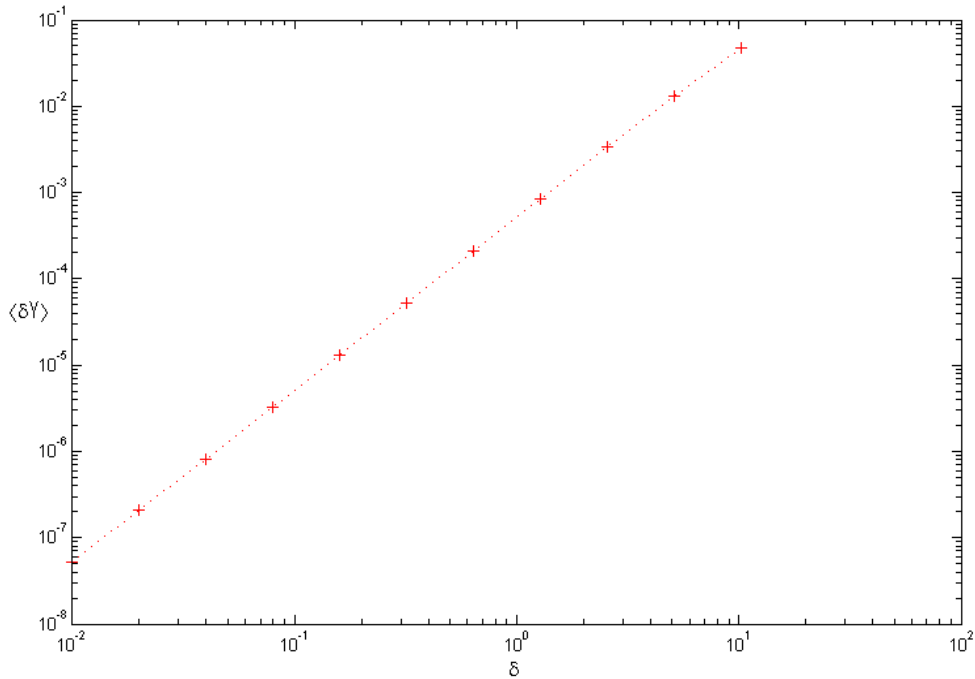

Fig.4- The average deviation of the yield as a function of $\delta$ for $N = 6, M = 4$

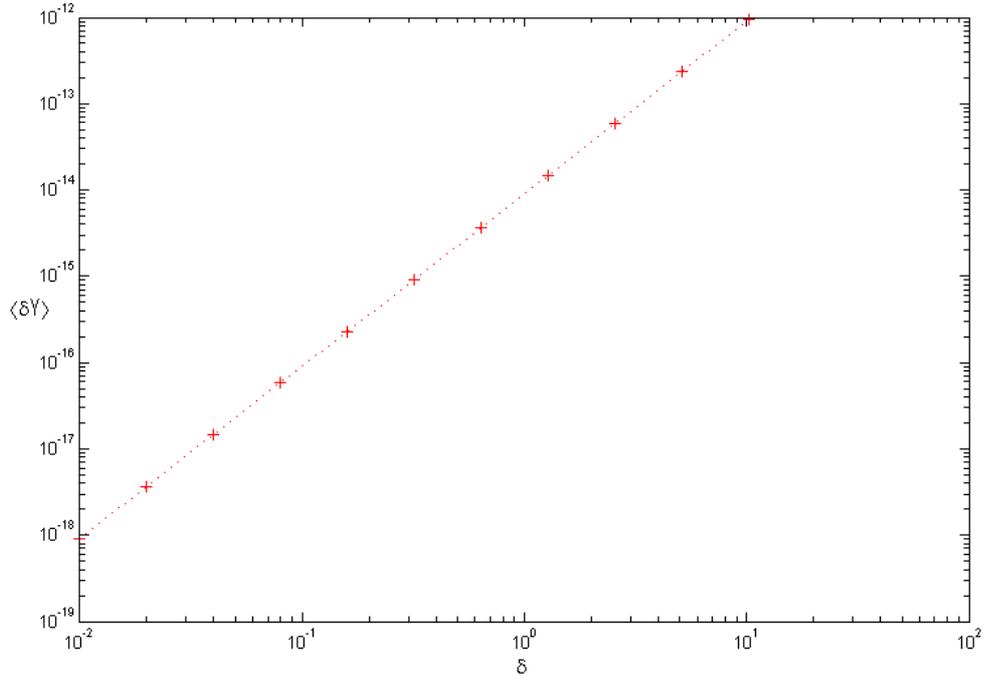

Fig.5- The average deviation of the yield as a function of $\delta$ for $N=12$, $M=8$

The relative deviation in the yield is given by

$$\frac{\delta Y}{Y} = \frac{\int_{-a}^{a} \tilde{f}(x)\delta(x)dx}{\int_{-a}^{a} \tilde{f}^2(x)dx} - 2\frac{\int_{-\pi}^{\pi} \tilde{f}(x)\delta(x)dx}{\int_{-\pi}^{\pi} \tilde{f}^2(x)dx}. \tag{14}$$

The error in the yield clearly obeys

$$\frac{|\delta Y|}{Y} < 2[\frac{\left|\int_{-a}^{a} \tilde{f}(x)\delta(x)dx\right|}{\int_{-a}^{a} \tilde{f}^2(x)dx} + \frac{\left|\int_{-\pi}^{\pi} \tilde{f}(x)\delta(x)dx\right|}{\int_{-\pi}^{\pi} \tilde{f}^2(x)dx} \tag{15}$$

and by the Schwartz inequality and equation (12) above,

$$\frac{|\delta Y|}{Y} < \frac{4\sqrt{N}\delta}{\pi}[\frac{a^{1/2}}{[\int_{-a}^{a} \tilde{f}^2(x)dx]^{1/2}} + \frac{\pi^{1/2}}{[\int_{-\pi}^{\pi} \tilde{f}^2(x)dx]^{1/2}}]. \tag{16}$$

The second term in the square brackets above can be neglected compared to the first and taking into account that within the superoscillating interval, $\tilde{f}^2(x)$ is of order 1, we arrive at the conclusion that

$$\frac{|\delta Y|}{Y} < \beta \frac{\sqrt{N}\delta}{a^{1/2}}, \qquad (17)$$

where $\beta$ is a constant of order 1.

Thus, if the condition for the preservation of superoscillation under introducing errors in the optimal Fourier coefficients is obeyed, it ensures automatically a limitation on the error in the yield. In the following we give the average relative error in the yield, which we call yield sensitivity.

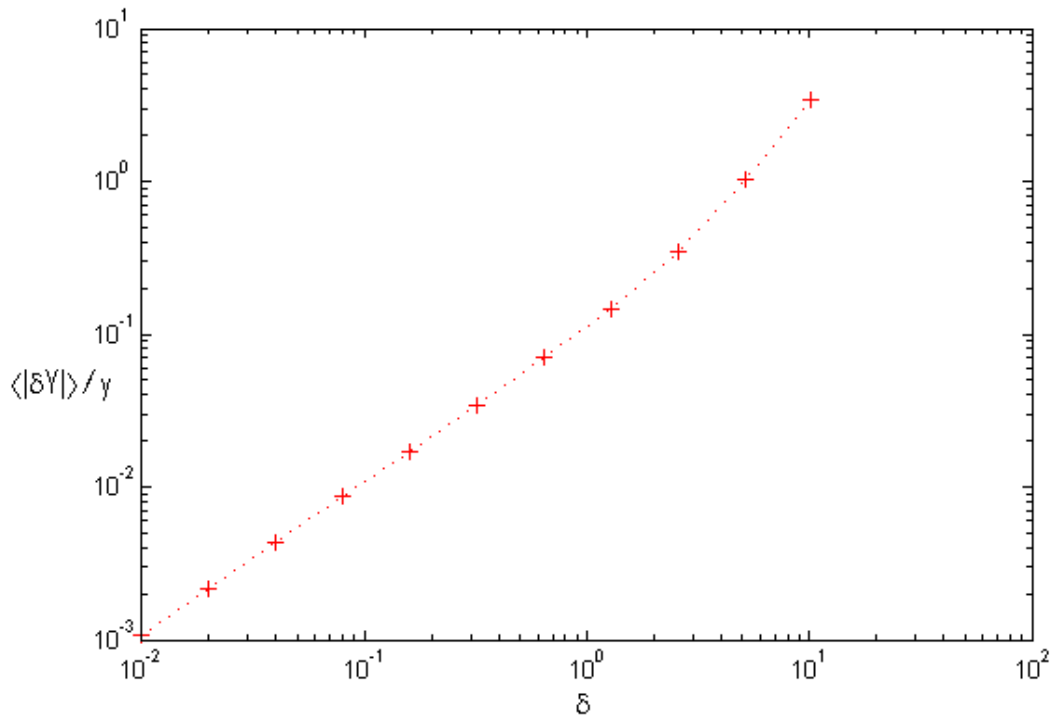

Fig. 6- Yield sensitivity for $N = 6$, $M = 4$.

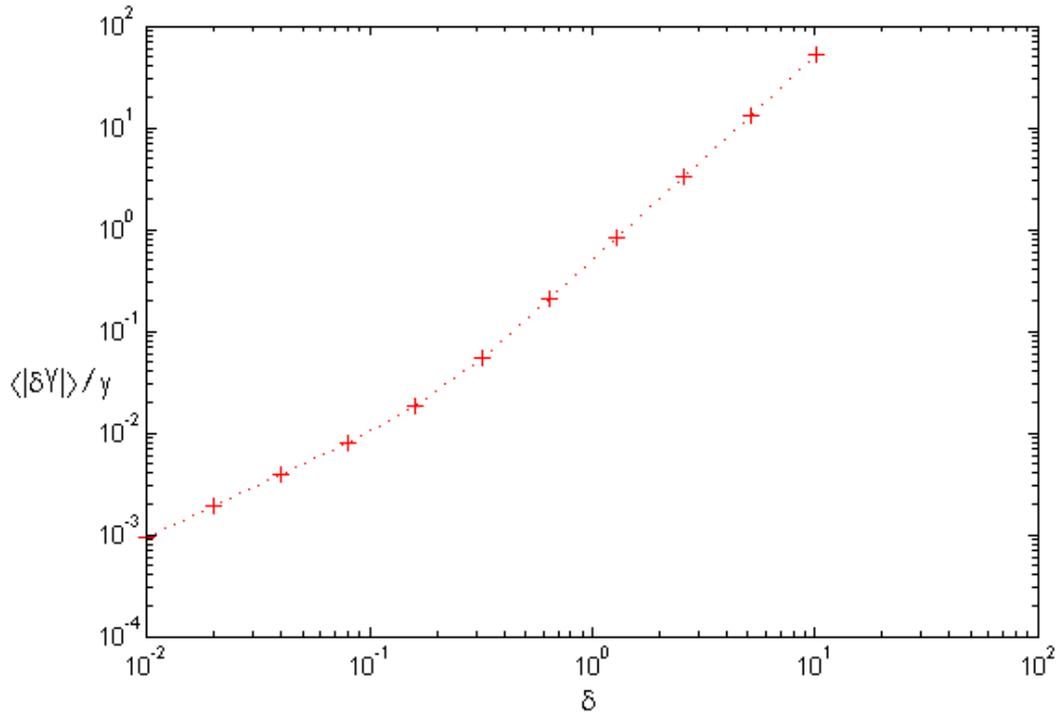

Fig. 7- Yield sensitivity for $N = 12$, $M = 8$.

Since small changes in $\delta$ result in large changes in the average of the deviation of the yield from its optimum or equivalently to large deviations in the yield sensitivity we may expect also the distribution of the yield to be very wide. The scale in the graphs describing those quantities suggest that if we would like to obtain a distribution, it would be reasonable to obtain the distribution of the logarithm of the yield rather than the distribution of the yield itself. That could result in a relatively narrow distribution. Furthermore, after obtaining the distribution of $\delta V$ and the distribution of the logarithm of the yield, it would be interesting to ask if somehow the deviation from the constraints is related to the deviation in the yield. It is natural to assume that the interpolation constraint reduces the yield considerably. Therefore we may expect that the larger $\delta V$ is the larger is the yield.

To check our first assumption we present first the distribution of the logarithm of the yield (figure 8).

To check our second assumption, we present in the following the scatter of $\delta V$ and $\delta Y$ for fixed $\delta$. Namely, we chose a thousand sets $\{\delta_m\}$ from the Gaussian distribution with standard deviation $\delta$, for each set we obtain $\delta V$ and $\delta Y$ and present the pair as a point in the $\delta V \delta Y$ plane. Both graphs are presented for $\delta = 0.01$, $N = 12$ and $M = 8$.

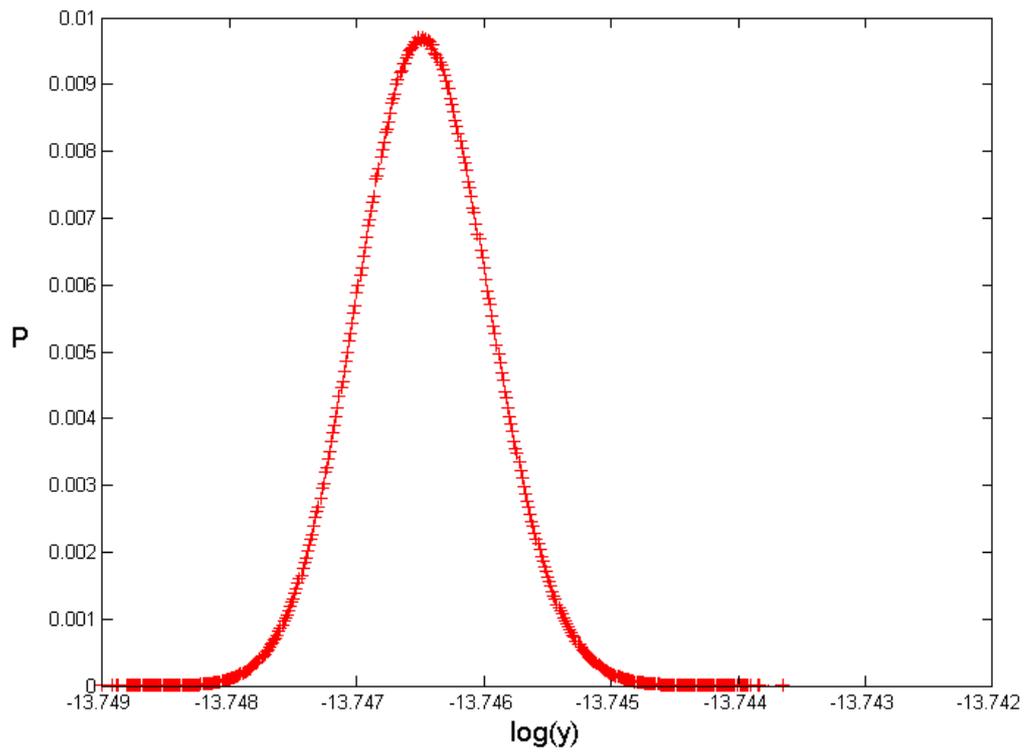

Fig.8- The normalized distribution of the logarithm of the yield, for $\delta = 0.01$.

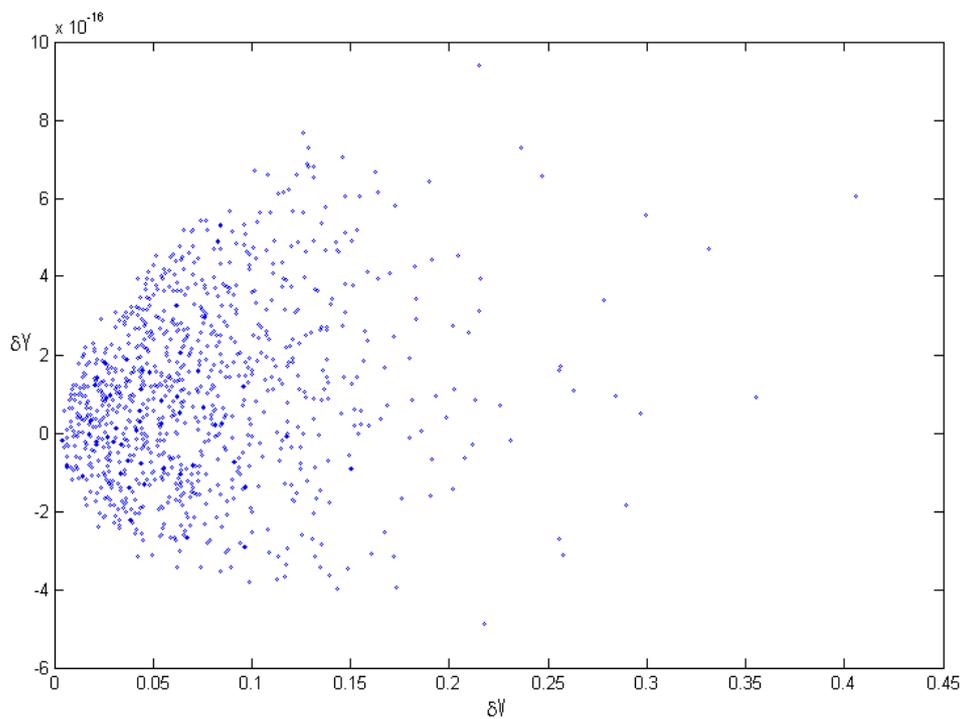

Fig. 9 The scatter of $(\delta V, \delta Y)$ for $\delta = 0.01$, $N = 12$ and $M = 8$.

We see first that indeed, the distribution of the logaithm of the yield is narrow and resembles a Gaussian. We see also that    when we increase $\delta V$ the tendency to obtain larger positive $\delta Y$'s increases considerably more than the tendency to obtain larger megative $\delta Y$'s. So, that the assumption that relaxing the constraints will yield higher yields seems to be correct. That is also the reason why $\langle \delta Y \rangle$ is always positive.

To conclude we have shown in this paper that errors in the Fourier coefficients of the low frequemcy components of the superoscillating signal affect both the interpolation sensitivity and the yield sensitivity. Estimates on both sensitiivities as a function of the number of the Fourier components and the standard deviation describing a single error have been obtained. We found that to ensure superoscillations at the desired frequency, may be more difficult if only a relative error  of a minimal size can be achieved. If an absolute minimal error can be achieved regardless of the size of the optimal Fourier coefficients, the superoscillations can be preserved easily under random errors in the Fourier coefficients. The latter condition is enough to ensure that the deviation of the yield from its optimal value (under the interpolation constraints) is relatively very small.